\documentclass[12pt,preprint]{aastex}

\newcommand\ms{M$_{\odot}$}

\slugcomment{The Astrophysical Journal, in press}

\shorttitle{Planetary Nebulae Abundances}
\shortauthors{Stanghellini et~al.}
  
\begin{document}
  
\title{Planetary Nebula Abundances and Morphology:
Probing the Chemical Evolution of the Milky
Way}
\author{Letizia Stanghellini}
\affil{National Optical Astronomy Observatory, 950 N. Cherry Av.,
Tucson, AZ  85719}
\email{lstanghellini@noao.edu}

\author{Mart\'in Antonio Guerrero}
\affil{Instituto de Astrof\'isica de Andaluc\'ia, Consejo Superior de 
Investigaciones Cient\'{\i}ficas, Apartado Correos 3004, 
E-18080 Granada, Spain}
\email{mar@iaa.es}

\author{Katia Cunha}
\affil{National Optical Astronomy Observatory, 950 N. Cherry Av.,
Tucson, AZ  85719}
\email{kcunha@noao.edu}

\author{Arturo Manchado}
\affil{
Instituto de Astrof\'isica de Canarias, v\'{\i}a L\'actea s/n, 
La Laguna, E-38200 Tenerife, Spain}
\email{amt@iac.es}

\author{Eva Villaver}
\affil{Space Telescope science Institute, 
3800 San Martin Drive, Baltimore, MD 21218}
\email{villaver@stsci.edu}

\clearpage
\begin{abstract}

This paper presents a homogeneous study of abundances in a sample of 79 northern galactic 
planetary nebulae whose morphological classes have been uniformly 
determined. Ionic abundances and plasma diagnostics
were derived from selected optical line strengths in the literature, and elemental 
abundances were estimated with the Ionization Correction Factor developed by Kingsbourgh \& Barlow (1994). 
We compare the
elemental abundances
to the final yields obtained from stellar evolution models of low-
and intermediate-mass stars, and we confirm that 
most Bipolar planetary nebulae have high nitrogen and helium abundance, and are the likely
progeny of stars
with main-sequence mass larger than 3 \ms.  We derive ${\rm <Ne/O>}$=0.27, and discuss the 
implication of such a high ratio in connection with the solar neon abundance. 
We determine the galactic gradients of oxygen and neon, and found
$\Delta$ log ${\rm (O/H)}$/$\Delta$~R=-0.01 dex kpc$^{-1}$ and $\Delta$ log ${\rm (Ne/H)}$/
$\Delta$~R=-0.01 dex kpc$^{-1}$. These flat PN gradients do not reconcile with
galactic metallicity gradients flattening with time. 

\end{abstract}


\section{Introduction}

Planetary nebulae (PNe) are ejected at the tip of the Asymptotic Giant Branch (AGB)
evolutionary phase in low and intermediate mass stars, and they have been used 
as probes of the galactic nucleosynthesis history and  stellar evolution 
in many studies. The PN abundances may be used as probes of primordial
nucleosynthesis when studying  the elements with zero yields through the AGB evolution
(oxygen, neon, and argon) and as probes of stellar evolution through the analysis of the 
products of the evolutionary nucleosynthesis
(helium, nitrogen, and carbon). 

One important aspect of PN chemical composition studies is the application to determining
of abundance gradients in the galaxy.
Metallicity gradients represent one of the most important
constraints for models of galaxy formation and evolution. Besides PNe, metallicity 
gradients can be measured in other galactic populations, ultimately
probing distinct time periods covering the formation and evolution of the galactic disk.
For the young populations, 
metallicity gradients are measured using H II regions and  OB stars;
the intermediate age populations are represented by Cepheid variables, while
the oldest populations are probed with studies of open clusters and PNe.
A question of significant importance in this context is whether there has been 
an increase or decrease of the abundance gradients with time.

In this study we present results from a homogeneous abundance analysis
for the elements helium, nitrogen, oxygen, neon, and argon in a large sample of
PNe 
and derive galactic metallicity gradients for the elements oxygen and neon.
The strength of the present study is multi-fold. First, our PNe 
are uniformly distributed within the galaxy, with a well-defined coverage
($0^o < l < 240^o$); second, the PN distances adopted here are the 
best available for each nebula; third, the abundances
were obtained homogeneously and self-consistently; and last, the morphology of these PNe 
has been classified in a uniform fashion, and used for stellar population selection. 
In $\S$2 we review the morphological classification that has been used, and the choice of PN distances. In $\S$3 
we describe the method used for abundance determination. In $\S$4 we present the results of our
analysis, and we summarize our findings and conclusions in $\S$ 5.

\section{Morphology and Distances}

The PN sample used in this study consists of galactic PNe in the northern 
sky that have a morphological determination. The PN images were selected mainly from the
Manchado et al. (1996) catalog, complemented by
those PNe from Balick's (1987) and Schwarz et al.'s (1992) samples needed to populate homogeneously
the volume of space defined by Manchado et al.'s survey, as described in
Manchado et al. (2000). 
The PN morphological
classes used here are described in Stanghellini et al.
(2002): Round (R), Elliptical (E), and Bipolar (B) PNe. Bipolar Core (BC) PNe, those
nebulae whose main contour is elliptical and that show inner ring structure, 
are included in the major elliptical class except when explicitly noted.
We used this database to explore the chemical evolution of the Milky
Way, using morphology as an indication of stellar population. 

Several recent studies have determined that PN morphology is a powerful indicator
of stellar population. In particular, R PNe may belong to the solar neighborhood
(Manchado et al. 2000), and may be the progeny of the lower mass progenitors
(Stanghellini et al. 2002). Elliptical PNe are found uniformly through the galactic disk,
thus they represent the low to intermediate mass progenitors. Bipolar PNe have typically
very low altitudes on the galactic plane, high nitrogen abundances, and possibly even high 
mass central stars (Stanghellini et al. 1993 and 2002, Corradi \& Schwarz 1995, Manchado et al. 2000).
These characteristics seem to conjure that the progenitors of B PNe belong to the higher-end
mass range of the AGB stars (M$>$3 \ms). Finally, BC PNe show a morphological complex 
or ring-like structure, but they do not have extended lobes. They are probably a mixed
group, with some E with inner structures and a few B PNe whose lobes are
below the surface brightness limits of the observations. From a population viewpoint they 
may belong either to the E or B groups.

Distances to galactic PNe are very hard to acquire. While precise individual distances, 
are known only for a small faction of galactic PNe, statistical distance scales are calibrated
to allow the derivation of an approximate distance for the rest of the galactic sample. 
The statistical method by Cahn et al. (1992) is based on the relation between the (inverse) surface
brightness and the ionized mass of the PNe. Other methods are based on similar relations that hold
for PNe on a physical basis, and all statistical distance scales are calibrated
with PNe of known distances. 
In Cahn et al. (1992) the maximum scatter of the distances is $\Delta$d/d $\sim$0.4.
As all statistical distance scales, the Cahn et al.
scheme is only accurate as long as one can measure unambiguously the area of the PN projection on the sky plane, 
and this is not straightforward for B PNe.
This fact should be taken into account when using bipolar PN and a statistical distance scale
to infer gradients and other distance-dependent quantities. 

In this paper we select
individual distances from Acker et al.~(1992) for the PNe where at least
one individual distance was available, and we use their averages in our gradient calculations. 
In all other cases, we use the
statistical distances from Cahn et al. (1992), where available, or those calculated with the same method
by Stanghellini et al. (2002). We believe that our distances, both individual 
averages and statistical, are the best possible
available for the PNe in our sample. 
We have compared our statistical distances to individual 
averages for the PNe whose individual distances are available, and found
a correlation coefficient of 0.6, which made us comfortable to use 
our statistical distances for the remaining PNe.

\section{Abundances}

In order to overcome the inhomogeneity problem that afflicts most PN abundance 
compilations we have compiled our own set of relative emission
line intensities for the PNe in our sample and analyzed them in a homogeneous fashion. 

Most of the relative emission line intensities have been adopted
from the {\it Catalog of Relative Emission Line Intensities Observed
in Planetary Nebulae} (Kaler, Shaw, \& Browning 1997), also available
on line, including all high-quality emission line fluxes published before May 1995. 
We complemented the Kaler et al. (1997) catalog with 
emission line fluxes from additional, more recent references
(Guerrero \& Manchado 1996 for A~58;
Kaler et al. 1996 for K~3-51, 
K~3-94, M~2-52, and M~3-28:
Trammell et al. 1993 for CRL~618:
Guerrero et al. 1996 for K~4-55; and 
Cuisinier et al. 1996 for PC~19).
In total, we have assembled a database of emission line
intensities (emission line intensities relative to H$\beta$)
for 19 R, 45  E (9 of which BC), and 15 B PNe. Overall, in this study 
we can count on 79 northern galactic
PNe with homogeneously derived abundances and morphologies.

The published intensities have then been averaged, uncorrecting
for extinction in the few cases where needed, by using the originally published extinction constants. 
We have used only emission intensities from the PN as a whole, and
excluded from averages these intensities relative to halos, knots, and 
fliers.
We then corrected the intensity ratio averages for interstellar
extinction, using the theoretical intensity ratios by Brocklehurst (1971),
and the standard reddening law by Whiteford (1958).

The plasma diagnostic ($N_{\rm e}, T_{\rm e}$) and ionic chemical
abundances have then been computed using the
nebular analysis package in IRAF/STSDAS (Shaw et al. 1998).
The helium abundances have been computed including the collisional
effects described by Clegg (1987).
Finally, the elemental abundances have been determined applying
the ionization correction factors (ICFs) following the prescription
that Kingsburgh \& Barlow (1994, hereafter KB94) give for the case where only optical lines are available.

In Table 1 we give the elemental abundances derived as described above. 
The first column gives the common PN name, then we give the helium, nitrogen, oxygen, neon, 
and argon abundances in terms of hydrogen, as X/H, where X represents the chemical
elements. Note that the helium abundance 
has been multiplied by $10^2$,
nitrogen, oxygen, and neon values have been multiplied by $10^4$, and argon abundances by $10^6$.
Averages of all elemental abundances for each morphological type are reported at the end of each morphological group in the 
Table. We estimate that typical uncertainties are $\sim10\%$ for the oxygen, helium, and
nitrogen abundances, and $15$ to $20\%$ for the neon and argon abundances, depending on the  
ionization stages available in the analyzed spectra.

We compare our abundances to large samples of abundances published in recent years. 
In Table 2 we report the average oxygen abundances and their standard deviation for
three PN samples: our own PNe (column [2]); the sample that we have in common with  
Perinotto et al. (2004, hereafter PMS04),
which includes 27 PNe (column [3]); and the sample that we have in common with Henry et al. (2004, hereafter HKB04), 
which includes
17 PNe (column [4]). 
In row 1 we give the averaged oxygen abundance from our work, for the three samples respectively.
In row 2 we give, in columns 3 and 4, the averaged oxygen abundances using the abundances from PMS04 and 
HKB04 respectively, and the samples in common with our sample, so that the columns are readily comparable.

Perinotto et al. (2004) calculated abundances from averages of 
published intensity lines. We have 27 objects in common with their sample, and our
average oxygen abundance (and its scatter) compare reasonably well with theirs (column 3, Table 2)
for the object in common.
By correlating the individual oxygen abundances of PNe in common between our sample 
and that of PMS04 we obtain a correlation coefficient of 0.5,
and most oxygen abundances in common agree within the uncertainties. 

Henry et al. (2004) have 
determined the abundance of a large sample of PNe from an homogeneous 
set of observations. Henry et al.'s work overlaps with our
morphological data-set with 17 PNe in common. Abundances of PNe from ours calculations and HKB04's
correlate well (with correlation coefficient of 0.8, 0.7, and $0.6$  for O/H, N/H, and Ne/H respectively)
but the HKB04 oxygen and neon abundances
are systematically higher (by a factor of $\sim 1.5$) than ours. We attribute this offset to the
use of different Ionization Correction Factors (ICF) for oxygen and neon. HKB04 used ICF(O)= (He$^+$ + He$^{++}$)/He$^+$, as
described in Kwitter and Henry (2001),
while we have used ICF(O)= ((He$^+$ + He$^{++}$)/He$^+$)$^{2/3}$,
as defined by KB94. 
The ICF for oxygen abundance is also
a scaling factor for the neon ICF, and this explains the scaling of neon abundances as well. 

Costa et al. (2004) measured oxygen abundances for a sample of anti-center
PNe, also used ICF(O)= (He$^+$ + He$^{++}$)/He$^+$. We have only six PNe in common with the Costa et al.'s
(2004)
sample, and we recover an offset between our and their abundances, their oxygen
abundances being $\sim 0.18$ dex higher than ours, confirming the origin of the abundance offset 
using different ICFs.
In order to use our neon and oxygen 
abundances together with another sample of published abundances
a re-scaling by the appropriate ICF ratio is advised, if using ICFs different than those of KB94.

\section{Results}

\subsection{Effects of Stellar Evolution}

In Figure 1 we plot the N/O ratio against helium abundances for the PNe
in our sample\footnote{We use the notation A(X)=log X/H +12, thus A(N)-A(O)=log (N/O).}.
This is a classical diagnostic plot to discriminate between Type I and
non-Type I PNe (Peimbert \& Torres-Peimbert 1983), and to study the connection
between high mass PN progenitors and enhanced chemical abundances of helium and nitrogen
(Kaler et al. 1990). 
We have plotted the PNe of our study with different
symbols for the different morphological types (hereafter 
open circles: R; asterisks: E; triangles: BC; squares: B PNe).
In Figure 1 we note the high segregation of B 
PNe toward the higher helium and N/O. 

The N/O ratio is a well-known diagnostic ratio for nucleosynthesis in AGB stars.
Nitrogen is produced in AGB stars in two ways: by neutron capture, during 
the CNO cycle, and by hot-bottom burning, if the base of the convective 
envelope
in AGB stars is hot enough to favor the conversion of $^{12}$C to $^{14}$N. 
The hot-bottom burning produces primary nitrogen, 
contrary to the CNO cycle, that produces secondary carbon and nitrogen.
Since hot-bottom burning may occur only in the most massive of the AGB stars
(with turnoff mass M$_{\rm TO}$ larger than  3 \ms, van der Hoek and Groenewegen 1997), nitrogen is expected to be
enriched, with respect to oxygen, in those PNe whose progenitors
were the most massive. A correlation between nitrogen enrichment and high mass
central stars has been observed by Kaler and Jacoby (1990).
The helium abundance also depends on the initial mass of the progenitor.
Helium is enriched progressively from M$_{\rm TO}$=1 to 3 \ms, 
reaches a plateau between 3 and 4 \ms, and then increases again
toward the higher masses (Marigo 2001). 

In this work, we have not defined Type I PNe {\it a priori}, but we 
obtain that morphologically distinct PNe belong to different loci 
in the plane of Fig.~1. 
By comparing the data distribution of Fig.~1 to
the stellar evolutionary models of Marigo (2001; thick line for M$_{\rm TO}> 3$\ms, thin line
for M$_{\rm TO}< 3$\ms) we see that all the B PNe are within the domain of the high mass stars,
while the R PNe are predominantly in the low-mass domain, while most E PNe
are in the same locus as the R PNe, a few in the B PN locus.

It is worth noting that the yields from theoretical models reflect very specific initial conditions
for the elements that do not vary with AGB evolution. For example, the range of A(O) in
Marigo's models (8.90 - 9.0) is much more narrow than the observed one (7.73 - 9.09).

\subsection{Oxygen, Neon, and Argon Abundances}

One remarkable result of our abundance analysis is the very tight correlation
between neon and oxygen abundances, as plotted in Figure 2. 
The correlation coefficient between the two abundances is 0.9 for 
the whole sample. We calculated the average ratio of the neon versus oxygen
abundances, obtaining $<$Ne/O$>$=0.27$\pm$0.10 for the whole sample, and,
consistently, $ <$Ne/O$>$=0.27$\pm$0.09 for E PNe alone. The almost identical
average of the Ne/O ratios for the whole sample and for 
Elliptical PNe indicates that neon and oxygen abundances are locked
independent of morphology, i.e.,
independent
on stellar population.
The relation between neon and oxygen abundances, and its independency on
PN population, is a direct consequence of the fact that both
neon and oxygen derive from primary nucleosynthesis, mostly
in stars with M$>$10~\ms, thus independent on the evolution of the PN
progenitors through the AGB. It is worth noting that the same correlation and averages would have been found 
even if we had used the alternative ICF (e.~g. HKB04).

We do not find a correlation between the argon and oxygen (or neon) abundances, 
with global correlation coefficient of 0.1. Argon and neon both derive from
primary nucleosynthesis, respectively from the carbon and oxygen burning,
occurring in Type I but especially Type II SNe. The nucleosynthesis of argon
necessitates the presence of alpha particles, and it is enhanced in presence of
secondary elements such as carbon and oxygen in the environment (Clayton 2003). This may
be the cause for the distinct trends of argon and neon in PNe, or, more likely, the
larger uncertainties in the argon abundance calculation hid a possible correlation between
the two species.

\subsubsection{Comparison with Solar Ne/O and Implications for Solar Models}

The average Ne/O ratio obtained from our PN sample is almost 
a factor of 2 higher than the most recent assessment of Ne/O in the Sun by Asplund et al.
(2005) that recommends a value of Ne/O=0.15. The solar Ne/O ratio, however,
is uncertain because the neon abundance in the Sun is not measured from lines that
form in the solar photosphere, but from high energy particles in the solar corona.
The average Ne/O ratio obtained from our sample is very consistent 
with an average Ne/O computed for the sample analyzed in HKB04 (<Ne/O>=0.28 for the
PNe in common with our sample).
These higher than solar Ne/O ratios obtained for PNe, 
both in this study as in HKB04, are interesting when put
into the context of the recent debate about the Ne/O ratio in the Sun and
implications for solar models. 

One of the successes in solar physics
has been the good agreement obtained between predictions from solar models and
oscillation measurements from helioseismology. Recently, however, the solar abundances
of key elements such as, C, N, and O have been revised downward by roughly 0.2-0.3 dex,
according to abundance analyses based upon more realistic 3-dimensional
hydrodynamical model atmospheres (see Asplund et al. 2005 for a review).
Solar models that adopt these revised solar abundances as input can no
longer satisfactorily match the helioseismological observations.
One way to reconcile the solar models with observations, however,
is to adopt a significantly higher neon abundance for the Sun.
This possibility has been investigated by Bachall et al.
(2005) who concluded that it would suffice to raise the neon abundance
in the Sun to 8.29 (or, Ne/O= 0.42) in order to bring solar models back into agreement
with the helioseismological measurements.  

A least-squares fit to the abundances obtained for the PN in this study 
yields the following relation for neon versus oxygen: (Ne/H) = 0.3446 (O/H)  - 2.348 $\times 10^{5}$.
If we plug in the solar oxygen abundance value of A(O)=8.67 (Allende Prieto,
Lambert, \& Asplund 2000) in this relation, we obtain A(Ne)=8.14, which is 
significantly higher than the currently estimated solar value of A(Ne)=7.84. 
We argue that the high Ne/O obtained for PNe provides
indirect support to a potentially higher neon abundance in the Sun, in comparison
with the recommended value by Asplund et al. (2005). This argument is based
on the simple fact that  a fraction of
the PN in our sample evolved from progenitor stars with low masses.
In particular, if we consider the Round PNe, which are
very likely the progeny of one solar mass stars, we obtain an average
Ne/O = 0.26. Therefore, the results obtained for PNe would be more
in line with a somewhat higher neon abundance that is needed in order to, at least partially,
solve the solar model problem.

\subsection{Galactic Metallicity Gradients}

In order to obtain the elemental gradients we calculate the
PN galactocentric distances with the standard formulation (e.~g., Maciel \&
Lago 2005). We used as input R$_{\odot}$=8.0 kpc, the distance to the galactic center,
and we used statistical and individual PN distances as described in $\S$2. 
Galactocentric distances used here are given in column (7) of Table 1, where we
indicate whether the distance is derived from Cahn et al.'s (1992) statistical scheme
(superscripts 1 to 4), or
is an average of individual distances from Acker et al.'s (1992) (superscript 5).  
We estimated the random uncertainty 
of the statistical distances by assuming the maximum relative error of 40 $\%$ in the heliocentric
distances from Cahn et al. (1992) (see $\S$2), and by assuming that the error in the 
galactic longitude and latitude of the PNe is very small in comparison with the distance error.
The final relative errors in the galactocentric distances are rarely higher than 60$\%$, 
the random uncertainty being 20 $\%$ on average. In Table 1 
we note -- with superscripts 1 to 4 -- whether the relative distance error is better than 20 $\%$, between 20 and 40 $\%$, between
40 and 60 $\%$, and larger than 60$\%$. These errors only take into account the (maximum)
errors from the
statistical distance scale, but not the asymmetry of the PN: for B PNe, the uncertainty on the
statistical distance may be larger than quoted, depending on the nebular shape.
The uncertainties of the 
individual distances in Table 1 can be found in the original papers, as quoted by Acker et al.
(1992). We calculate the spread of individual distances, where more than one 
distance was available, and found that the different determinations agree to 
better than 10 $\%$  in most cases, with an average spread of about
5 $\%$ in the individual distance determinations.
Note that individual distances and their averages do not suffer the B PN uncertainty, since they are
not based on angular diameter measurements.
We did not use halo PNe in the gradient analysis, thus did not enter galactocentric distances 
for halo PNe in Table 1.

In Figure 3 we show the galactic gradients 
of the oxygen and neon abundances in the PNe of our sample.
In these figures we plot the different morphologies with distinct 
symbols, as described in Figure 1. We can study these gradients with the added
insight of the morphological classes. 

Bipolar PNe are the progeny of more massive stars that formed more recently, and
since oxygen and neon
abundance should not vary during the evolution through AGB and post-AGB, their oxygen
and neon abundances are probes of relatively young stellar population compared with R and
E PNe. A possible exception is the alternative pathway to neon production in low metallicity
2-4 \ms ~ stars (Gibson et al. 2005, and references therein).
Unfortunately B PNe are not very 
useful to establish galactic gradients for two reasons: first, they can not be seen at large
distances, due to their proximity to the galactic plane and consequent higher extinction. In Figure 3, for example, 
B PNe are probes of galactic gradients only between 3 and 9 kpc. Second, and most importantly, their statistical
distances may also be more uncertain than those of other PNe, since their 
sizes are not very well defined from the observations. 
Round PNe may belong to a faint, local population (Manchado et al. 2000). If this is true, their loci on
the plots of Figure 3 are deceiving, and they are poor probes of gradients as well.
It is best to use E and BC PNe for gradient estimates. 
Ideally, a gradient should be estimated with individual distances only. In out case, about 
1/3 of the PNe in our sample have individual distances, thus it represent an improvement with
respect to other samples. On the other hand, individual distances to PNe are only available for nearby objects,
thus have limited use for galactic gradients. In our sample, individual galactocentric distances are
in the 7--9 kpc range.

The oxygen gradient that we obtained including E and BC PNe of Figure 3 (top panel) is 
very shallow ($\Delta$ log $({\rm O/H)}$/$\Delta$~R = -0.01 $\pm$ 0.02 dex kpc$^{-1}$). 
The neon gradient obtained for the sample of E and BC PNe is the same than the oxygen 
gradient, as expected from the previous discussion of $\S$4.2.
PNe NGC 2242 and M2-44 have not been included in this gradient, 
due to large uncertainties in their neon abundances.

The oxygen and neon gradients determined by using the complete PN sample are identical to those
above, except that the scatter is reduced to 0.01 dex kpc$^{-1}$. The scatter in the gradients 
incorporate the maximum distance scale uncertainties and the abundance errors.

The argon 
abundance distribution has a slope of $\Delta$ log ${\rm (Ar/H)}$/$\Delta$~R=-0.05 dex kpc$^{-1}$ 
for the E and BC PNe,
with very large scatter
due to the larger uncertainties in the abundance derivation.

\subsubsection{Comparisons with Gradients from the Literature}

Several studies in the literature are devoted
to determinations of abundance gradients in selected samples belonging
to distinct populations in the galaxy. As discussed in the introduction, 
these include H II regions, 
young stars, Cepheids, open clusters and PNe. The abundances  
that define such gradients are derived from a diverse sets of objects that require 
very different analysis techniques. This diversity results in different systematic
errors in the abundances for the different populations of objects.  
In addition, measurements of metallicity gradients in the galaxy are uncertain due to
significant errors in the adopted distances for the individual objects. 
As a consequence, the magnitudes of
the best-fit slopes to the different data-sets of abundances versus galactocentric
distances have varied significantly, making in hard to separate the 
astrophysical variance (due to gradient changes with time, for example) from the
uncertainties. 

Compilations of elemental gradients obtained for galactic populations 
can be found in different studies (e.g. Matteucci 2003; Stasinska 2004).
In the following we mention results from the most recent studies of
gradients in young populations. For early-type stars, Daflon \& Cunha
(2004) conducted a homogeneous and self-consistent study of a large
sample of OB stars and found much flatter gradients than previous
studies of abundances in early-type stars. 
For oxygen, in particular, they found a relatively flat gradient of -0.03 dex kpc$^{-1}$
(neon and argon gradients have not been derived in studies of OB-type stars).
For H II regions, Deharveng et al. (2000) and Pilyugin et al. (2003) 
also found, for oxygen, a flatter gradient  
($\Delta$ log ${\rm (O/H)}$/$\Delta$~R=-0.04 dex kpc$^{-1}$) than previous H II region studies.
In addition, the gradients found recently for neon and argon by Martin-Hernandez et al.~(2002) are
also relatively flat: $\Delta$ log ${\rm (Ne/H)}$/$\Delta$~R=-0.04 dex kpc$^{-1}$ and 
$\Delta$ log ${\rm (Ar/H)}$/$\Delta$~R=-0.05 dex kpc$^{-1}$.

A recent determination of metallicity gradients based on PNe comes from
the work by HKB04. 
In Table 3 we assemble our gradients for the studied elements oxygen, neon and argon,
as well as other PN results from the literature. 
It is interesting to note that
the abundance gradients derived in HKB04 are slightly flatter than those obtained in 
Martins \& Viegas (2000, MV00) and Maciel \& Quireza (1999, MQ99). 
Our oxygen gradient is $\Delta$ log ${\rm (O/H)}$/$\Delta$~R=-0.01 dex kpc$^{-1}$, quite 
different from the steep value obtained previously by MQ99
(from abundances compiled from the literature);  
and still significantly flatter than the O/H gradient derived in HKB04. 
For neon, the situation is similar: our derived gradient is also much flatter than all
other studies. 

It is worth noting that all distances used by MQ99, as
well as by HKB04, 
derive form Maciel (1984) calculation, and those do not compare as well as
our statistical distances to individual distances. Interestingly, if we estimate the galactic
gradients from the abundances given in Table 1 and Maciel (1984) distances, we find 
$\Delta$ log ${\rm (O/H)}$/$\Delta$~R=-0.04 dex kpc$^{-1}$
and $\Delta$ log ${\rm (Ne/H)}$/$\Delta$~R=-0.03 dex kpc$^{-1}$, much closer to HKB04 or MQ99
gradients. It is natural to conclude that
the choice of distances can make a difference when estimating galactic gradients from
PN abundances.

Our O/H and Ne/H gradients have similar magnitudes: $\sim$-0.01 dex kpc$^{-1}$,
and carry average scatter of about 1 dex.
An agreement between oxygen and neon gradients is to be expected from predictions
of stellar nucleosynthesis. In principle, the oxygen and neon abundances 
evolve in lock-step as both of these elements are produced in short lived massive stars.
Agreement between neon and oxygen gradients, within the uncertainties, was also obtained for the
sample of galactic PNe studied by HKB04. A similar
result was derived by Crockett et al. (2006) from a study of
H II regions in M33. We note, however, that MQ99 do
not find agreement between the best-fit slopes for their neon and oxygen PNe data. 
For argon, our best-fit slope is steeper than the one obtained by HKB04,
but it is in agreement with MQ99. 
The Ar abundances, however, are
more uncertain and the obtained abundances scatter over a $\sim$ 1.5 dex interval at roughly 
galactocentric distances between 6 and 12 kpc. 

It is important to recognize that metallicity gradients derived for
populations of different ages are in fact probing distinct epochs of disk evolution.
In this context, Maciel et al. (2003) divided their PN sample in 3 
age groups so that possible time variations in the galactic gradients could
be investigated.
Maciel et al. (2003, 2005) concluded both from their PN sample, from open clusters samples 
(Friel et al. 2002; Chen et al. 2003)
and from comparisons with gradients obtained recently for young stars 
(Daflon \& Cunha 2004), and HII regions (Deharveng et al. 2000),
that there is significant flattening of metallicity gradients with time. 
The oxygen and neon gradients obtained in this study are flat and 
very difficult to reconcile with this picture, given that our sample of PNe 
do not support the existence of steep gradients earlier in the galaxy.

We also determined the gradients for two groups of PNe with low (log N/O$<$-0.5) and high
(logN/O$>$-0.5) nitrogen abundance, in order to reproduce the Type I and non-Type I gradients,
and found no significant differences between the two types, confirming that our results are
not compatible with gradients flattening with time.

Costa et al. (2004) have measured abundances of PNe in the direction of the galactic anti-center,
with galactocentric distances between 8 to 14 kpc.
Although they do not publish the gradient derived by their sample alone,
we calculate that their oxygen gradient is $\Delta$ log ${\rm (O/H)}$/$\Delta$~R=-0.01 dex kpc$^{-1}$,
which is also very flat and agrees with our gradient. 

For argon we find, unlike neon and oxygen, a steep gradient. As discussed
previously, however, the Ar abundances derived here 
are the most uncertain and there is a very large scatter in the obtained abundances.
In addition, argon abundances and respective gradients are measured in PNe,
but not in early-type stars, nor open clusters.
Therefore, it is not straightforward to
use this element in order to investigate time variations in the gradients.

\section{Conclusions}

In this paper we analyze elemental abundances of helium, oxygen, nitrogen, neon, and
argon for a sample of northern galactic PNe from the Manchado et al. (1996) morphological catalog,
complemented with other PNe in the Schwarz et al.'s (1992) and Balick's (1987) compilation that fit our
criteria, to build a volume complete sample. The
difference between this and previous studies is that we have used PN morphology as an extra 
tool to differentiate between the progenitors populations. Our abundances correlate well with previously published
samples that were calculated with similar ICFs. The abundance offset between ours and 
HKB04 (and Costa et al. 2004) oxygen and neon abundances is mostly due to these ICF differences.
Future studies should take these differences into account, and work on the comparison between the
HKB04 and other abundance determination methods is in progress (Henry and Kwitter, private communication).

We confirm that B PNe have higher N/O ratios and helium abundances than R and E PNe, marking the 
different range of progenitor's initial mass of the B
PNe with respect to R and E PNe.

We found a tight correlation between A(O) and A(Ne), and determined an average ratio of $<$Ne/O$>$=0.27$\pm0.10$.
We also discussed the implications that such a ratio has on the determination of solar abundances, and review the
possible alternatives.

By using the E and BC PNe, whose distances are reliable and whose
distribution is more uniform in the galaxy
we found gradients of the order of 
$\Delta$ log ${\rm (O/H)}$/$\Delta$~R=-0.01 dex kpc$^{-1}$ compared to previously determined gradients of
-0.04 dex kpc$^{-1}$. Our gradients are not consistent with a flattening of gradients with time.
Our abundances and distances are sound, and the abundances compare well with similarly determined
quantities by other groups. We conclude that the galactic abundance gradients are flat, but also that 
a definite answer will only be possible when more objects at larger and very small galactocentric 
distances will be detected, since the set of distances used for gradients have a great importance 
in the results. By selecting PN population in the inner disk one should be careful not to include bulge PNe, 
belonging to a different population. A different choice of the ICFs, while changing the elemental abundances, would
not affect the gradients. Furthermore,
the uncertainties in the statistical distances may affect the gradients, and more sets of individual distances
would improve the determination. Other classes of objects should be studied in addition of the PNe to extend the 
gradients to larger distances.

Given that PN distances carry such large uncertainties, we use distance-independent ways to
disclose abundance gradients. For example, if we average the oxygen abundance for PNe with 
$0^o<l<=45^o$ and with $135^o<l<240^o$
we sample respectively the PNe in the direction of the galactic center (GC) and anti-center (GA). We obtain 
$<$O/H$>_{\rm GC}=3.79\pm 2.26\times 10^{-4}$ and $<$O/H$>_{\rm GA}=2.91\pm 1.23\times 10^{-4}$, translating into
a difference of 0.11 dex in oxygen abundance 
between the galactic center and the anti-center directions, confirming the rather flat 
gradient that we derive in the more traditional way in this paper.

\acknowledgments

We warmly thank Karen Kwitter, Richard Henry, and Bruce Balick for thoroughly
discussing
with us the different abundance determination
methods, and Verne Smith for scientific discussion. We thank an anonymous Referee
for several important suggestion that improved our presentation.

\clearpage

\begin{deluxetable}{lcrrrrr}
\tablewidth{14truecm}
\tablecaption{Abundances and distances}

\tablehead{
\colhead{Name}&
\colhead{He/H} &  
\colhead{N/H} &  
\colhead{O/H}&    
\colhead{Ne/H}&   
\colhead{Ar/H}&
\colhead{R}\\

\colhead{}&
\colhead{$[\times 10^2$]} &  
\colhead{$[\times 10^4$]} &  
\colhead{$[\times 10^4$]}&    
\colhead{$[\times 10^4$]}&   
\colhead{$[\times 10^6$]}&
\colhead{[kpc]} \\
}

\startdata

\hline
{\it Round}\\
\hline

        A~4  &  8.2  &  2.36  &  4.26  & $\dots$  &  $\dots$& 12.8$^2$ \\
       A~50  &  3.8  &  1.20  &  3.30  & $\dots$  &  $\dots$& 8.5$^1$ \\
     Cn~3-1  &  5.7  &  0.58  &  2.16  &  0.23  &   0.16& 5.7$^2$  \\
     He~1-5  & 10.1  &  1.10  &  6.14  &  2.21  &  $\dots$& 7.2$^5$ \\
    IC~4593  & 12.7  &  0.36  &  2.94  &  0.59  &  $\dots$& 7.4$^5$  \\
      K~1-7  & 10.0  &  7.70  &  3.40  &  1.09  &  $\dots$& 13.4$^2$  \\
     K~3-27  & 11.0  & $\dots$  &  0.90  &  0.16  &   2.57& 7.0$^1$  \\
     K~3-51  & 10.5  & $\dots$  &  1.00  &  0.18  &  $\dots$& 6.7$^1$  \\
     K~3-73  &  7.9  &  0.35  &  1.70  &  0.77 &  $\dots$& 10.1$^3$  \\
     M~1-80  & 13.4  &  1.70  &  3.00  &  0.19  &   0.05& 11.0$^2$ \\
     M~4-18  &  7.7  &  0.71  &  3.30  & $\dots$  &  $\dots$& 14.5$^2$ \\
   NGC~2242  & 11.7  & $\dots$  & $\dots$  &  0.04  &   0.82& $\dots$  \\
   NGC~2438  & 10.3  &  0.85  &  4.50  &  1.03  &  $\dots$& 9.4$^5$ \\
   NGC~3587  &  9.9  &  0.29  &  2.30  &  0.90  &  $\dots$& 8.2$^5$  \\
   NGC~6879  & 11.4  & $\dots$  & $\dots$  &  0.69  &  $\dots$& 7.4$^2$  \\
   NGC~6884  & 13.5  &  2.10  &  4.54  &  0.82  &   2.18& 8.0$^5$   \\
   NGC~6894  & 11.1  &  4.49  & 12.35  &  5.16  &  $\dots$& 7.6$^5$   \\
       Na~1  & $\dots$  & $\dots$  & $\dots$  &  0.92  &   1.81& 3.5$^4$  \\
     Vy~2-3  & 11.3  &  0.15  &  2.00  &  0.64  &   1.43& 15.8$^3$  \\
     \hline
     
  Average &   10.0& 1.71& 3.61& 0.98& 1.29& \\

\hline
&&&&&\\
&&&&&\\
{\it Elliptical}\\
\hline

       A~2  &  7.2 &  1.00  &  3.38  &  1.18  &   6.40& 10.6$^2$  \\
       A~43  & 10.8  & $\dots$  & $\dots$  & $\dots$  &   4.05& 6.8$^1$ \\
       A~70  &  9.4  &  1.78  &  3.20  &  0.77  &  $\dots$& 6.5$^1$  \\
    CRL~618  &  4.9  &  0.56  &  3.40  & $\dots$  &  $\dots$& 11.4$^2$  \\
     Hu~1-1  & 10.9  &  1.50  &  5.80  &  1.96  &   0.30& 10.8$^2$  \\
     IC~351  & 11.1  & $\dots$  & $\dots$  &  0.38  &   0.89& 13.2$^2$  \\
       IC~2003  & 11.6  &  1.90  &  4.00  &  0.80  &   1.09& 10.6$^1$  \\
    IC~2149  &  9.6  &  0.29  &  1.61  &  0.16  &   0.65&8.9$^5$  \\
      J~320  & 11.4  &  0.32  &  1.51  &  0.46  &   0.62& 13.7$^2$  \\
     K~3-61  & $\dots$  &  7.83  &  3.84  &  0.70  &   0.28& 11.4$^2$  \\
     K~3-92  & 11.0  &  1.45  &  6.41  &  2.61  &  $\dots$& 13.4$^2$  \\
      M~1-6  & 11.5  &  0.07  &  0.54  & $\dots$  &   0.38& 10.4$^1$  \\
     M~1-64  & 11.3  &  1.70  &  2.60  &  1.10  &  $\dots$& 7.5$^1$  \\
      M~2-2  & 10.7  &  0.29  &  2.20  &  0.50  &   0.45& 11.9$^2$ \\
     M~2-44  & 29.1  & $\dots$  & $\dots$  &  0.05  &   1.63& $\dots$  \\
     M~2-50  &  3.9  &  1.10  &  2.30  &  0.55  &   1.37& 13.6$^3$  \\
      M~3-3  & 12.2  &  6.56  &  4.34  &  1.45  &   0.49& 12.9$^2$  \\
     Me~1-1  & 11.4  &  1.47  &  2.96  &  1.01  &   2.77& 6.4$^1$  \\
       NGC~40  &  9.3  &  0.79  &  3.30  &  0.26  &  $\dots$& 8.6$^5$  \\
        NGC~1514  & $\dots$  & $\dots$  &  1.10  &  0.37  &   1.32& 8.7$^5$  \\
   NGC~2022  & 13.4  &  0.88  &  4.50  &  1.06  &   2.03& 9.5$^5$  \\
   NGC~2392  & 11.2  &  1.14  &  2.36  &  0.51  &  $\dots$& 8.6$^5$  \\
   NGC~6210  & 11.4  &  0.50  &  3.60  &  0.90  &   0.66& 6.9$^1$ \\
   NGC~6543  & 11.2  &  0.92  &  4.59  &  1.39  &   0.65& 8.1$^5$  \\
   NGC~6720  & 11.7  &  1.80  &  6.00  &  1.70  &   0.32& 7.7$^1$  \\
   NGC~6741  & 23.1  &  5.90  &  4.40  &  1.02  &   2.61& 6.7$^5$  \\
   NGC~6826  & 10.6  &  0.23  &  2.52  &  0.52  &   0.30& 8.0$^5$  \\
   NGC~6853  & 12.1  &  2.20  &  7.00  &  2.24  &   1.88& 7.9$^5$  \\
   NGC~6905  & 12.8  &  0.80  &  5.00  &  1.45  &   0.85& 7.3$^1$  \\
   NGC~7008  & 12.7  &  0.47  &  1.47  &  0.77  &  $\dots$& 8.1$^5$  \\
   NGC~7048  & 13.5  & $\dots$  & $\dots$  & $\dots$  &  $\dots$& 8.1$^5$ \\
   NGC~7354  & 13.7  &  0.62  &  2.17  &  0.57  &   0.80& 8.9$^5$  \\
   NGC~7662  & 12.5  &  0.72  &  2.60  &  0.53  &   1.22& 8.3$^5$  \\
      PC~19  & 11.3  &  0.37  &  3.09  & $\dots$  &  $\dots$& 4.4$^2$  \\
     Vy~1-1  & $\dots$  &  0.76  &  2.30  &  0.46  &   0.56& 12.1$^2$  \\
     Vy~1-2  & 17.1  &  0.79  &  2.90  &  0.65  &   0.69& 7.6$^3$  \\
     
     \hline
     
Average&	12.0& 1.51& 3.34& 0.91& 1.31& \\

\hline
&&&&&\\
&&&&&\\
{\it Bipolar}\\
\hline
  
     K~3-46  & 10.3  & $\dots$  & $\dots$  & $\dots$  &  $\dots$& 7.5$^1$ \\
     K~3-94  & 13.7  & $\dots$  & $\dots$  & $\dots$  &  $\dots$& 13.7$^2$  \\
     K~4-55  & 14.0  & 10.00  &  2.40  &  0.54  &   3.30& 8.2$^1$ \\
     M~1-57  & 12.8  &  1.79  &  4.62  & $\dots$  &  $\dots$& 7.5$^3$  \\
     M~1-59  & 27.2  & $\dots$  & $\dots$  &  1.18  &  $\dots$& 6.9$^5$  \\
     M~1-75  & 21.0  & 23.00  &  5.30  &  1.57  &   3.88& 7.4$^1$  \\
     M~2-46  &  8.1  &  2.17  &  1.58  & $\dots$  &  $\dots$& 5.8$^4$  \\
     M~2-52  & 17.2  &  5.26  &  3.95  & $\dots$  &   1.03& 10.0$^2$  \\
     M~3-28  & 11.4  &  8.52  &  5.35  &  1.11  &  $\dots$& 3.9$^4$  \\
   NGC~2346  & 13.5  &  2.00  &  3.20  &  0.99  &   0.35& 8.9$^5$ \\
   NGC~6778  & 16.2  &  2.16  &  1.29  &  0.42  &   0.24& 5.7$^2$  \\
   NGC~6881  & 12.4  &  2.49  &  4.02  &  1.18  &   0.98& 7.7$^1$  \\
   NGC~7026  & 12.4  &  5.40  &  4.70  &  1.13  &   0.64& 8.2$^5$  \\
   NGC~7027  & 16.3  &  2.70  &  2.70  &  0.47  &   0.94& 8.0$^5$  \\
    Pe~1-17  & 17.6  &  6.10  &  9.46  & $\dots$  &  $\dots$& 3.4$^2$  \\
    
\hline
Average&	14.9&	5.97&  4.05& 0.95& 1.42& \\  
  
\hline
&&&&&\\
&&&&&\\
{\it Bipolar Core}\\
\hline
  
       He~1-6  & 12.8  & $\dots$  & $\dots$  & $\dots$  &  $\dots$& 7.3$^1$  \\
     Hu~1-2  & 18.5 &  1.90  &  1.90  &  0.52  &   0.86& 8.1$^5$  \\
      M~1-7  &  9.7  &  1.50  &  2.90  &  0.94  &   0.24& 13.8$^2$  \\
     M~1-79  & 13.2  &  1.77  &  1.59  & $\dots$  &  $\dots$& 8.4$^5$  \\
     M~2-51  & 14.8  &  1.98  &  3.80  & $\dots$  &  $\dots$& 8.6$^1$  \\
     M~2-53  & 14.9  &  1.63  &  4.43  &  1.93  &   0.56& 9.6$^1$ \\
     M~2-55  & 14.0  &  2.65  &  3.25  &  1.08  &  $\dots$& 9.2$^1$  \\
   NGC~6058  & 12.4  & $\dots$  & $\dots$  & $\dots$  &  $\dots$& 7.3$^1$ \\
   NGC~6804  & 15.0  & $\dots$  &  5.34  &  2.35  &   0.75& 7.0$^5$  \\
 
 \hline

 Average& 13.9& 1.91& 3.32& 1.36& 6.03& \\    
 
  \\
  \\

\enddata

$^1$ $\Delta$ R / R $<$0.2;
$^2$ 0.2 $< \Delta$R / R $<$0.4;
$^3$ 0.4 $< \Delta$ R / R $<$0.6;
$^4$ $\Delta$ R / R $>$0.6 \\
$^5$ Individual distances from Acker et al. 1992

\end{deluxetable}

\clearpage

\begin{deluxetable}{lrrr}
\tablewidth{14truecm}
\tablecaption{Oxygen abundance comparison}
\tablehead{
\colhead{$<$O/H$>$ $\pm$  $\sigma$} & 
\colhead{Our sample}&
\colhead{PMS04 sample}&
\colhead{HKB04 sample}\\
\colhead{[$\times$10$^4$]}& 
\colhead{67 PNe}&
\colhead{27 PNe}&
\colhead{17 PNe}\\
}

\startdata

This paper&	3.5$\pm$2.0&		3.9$\pm$2.2&  3.7$\pm$1.3\\

Other studies	&		$\dots$&		4.3$\pm$1.9&	5.6$\pm$1.7\\
		
\enddata
\end{deluxetable}

\begin{deluxetable}{lrrrr}
\tablewidth{12truecm}
\tablecaption{Gradients}

\tablehead{
\colhead{Ratio}&
\colhead{This work} &
\colhead{HKB04} &
\colhead{MV00} &
\colhead{MQ99}\\

\colhead{}&
\colhead{[dex kpc$^{-1}$]} &
\colhead{[dex kpc$^{-1}$]} &
\colhead{[dex kpc$^{-1}$]} &
\colhead{[dex kpc$^{-1}$]}\\}

\startdata

 O/H & -0.01 & -0.04 & -0.05  &  -0.06 \\
 Ne/H & -0.01 & -0.04 & -0.07 & -0.04  \\
 Ar/H & -0.05 & -0.03 & \dots     &  -0.05  \\

\enddata
\end{deluxetable}

\clearpage
\
\begin{figure}

\plotone{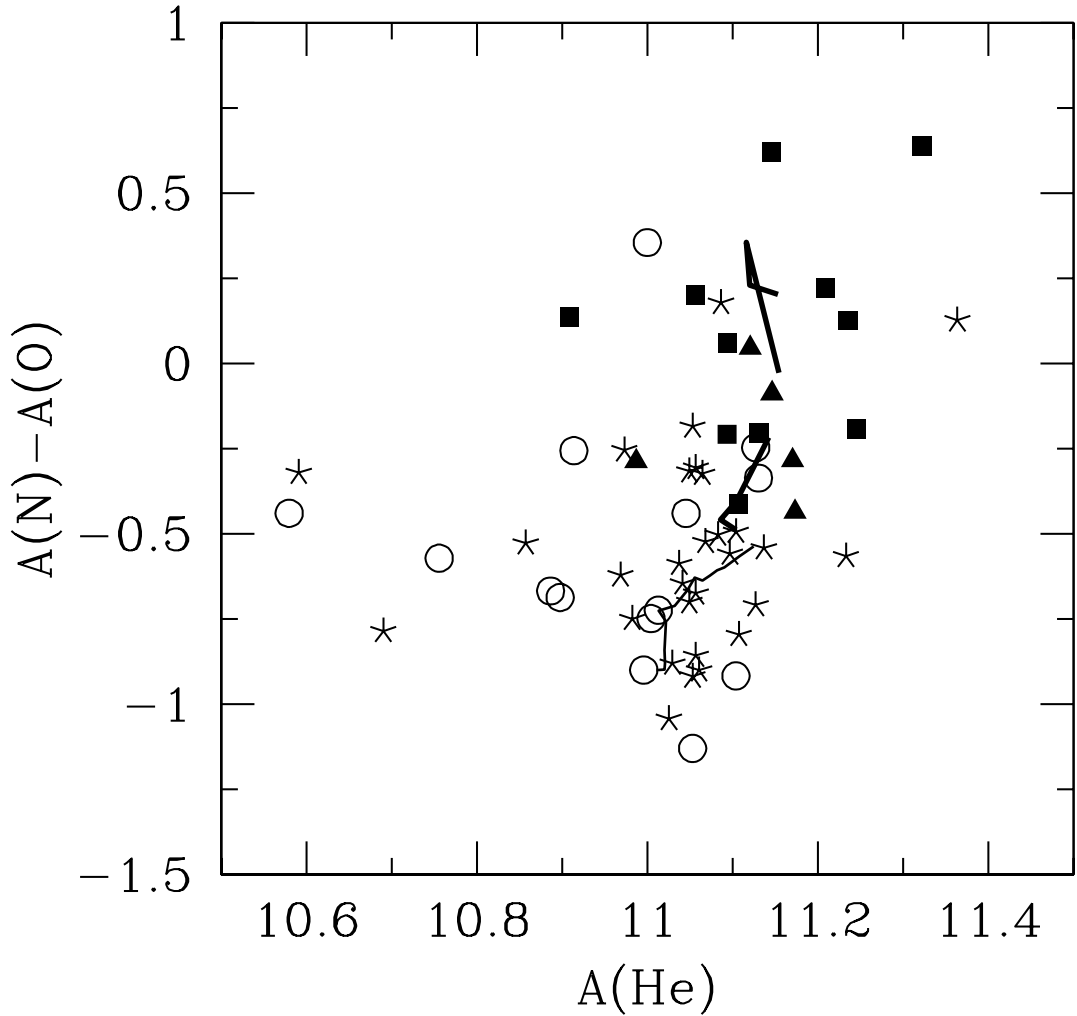}

\caption{A(N)-A(O), or log (N/O), versus A(He), or log (He/H)+12. 
Symbols indicate morphology types: Round (open circles), Elliptical 
(asterisks), Bipolar Core (triangles) and Bipolar (squares).
Solid lines: galactic models by Marigo (2001), thin line 
connect models with M$_{\rm TO}<3$~\ms~ and $\alpha$ -- the mixing-length parameter-- =1.68, thick line connects models
with M$_{\rm TO}>3$~\ms~ and $\alpha=2.5$. }

\end{figure}

\begin{figure}

\plotone{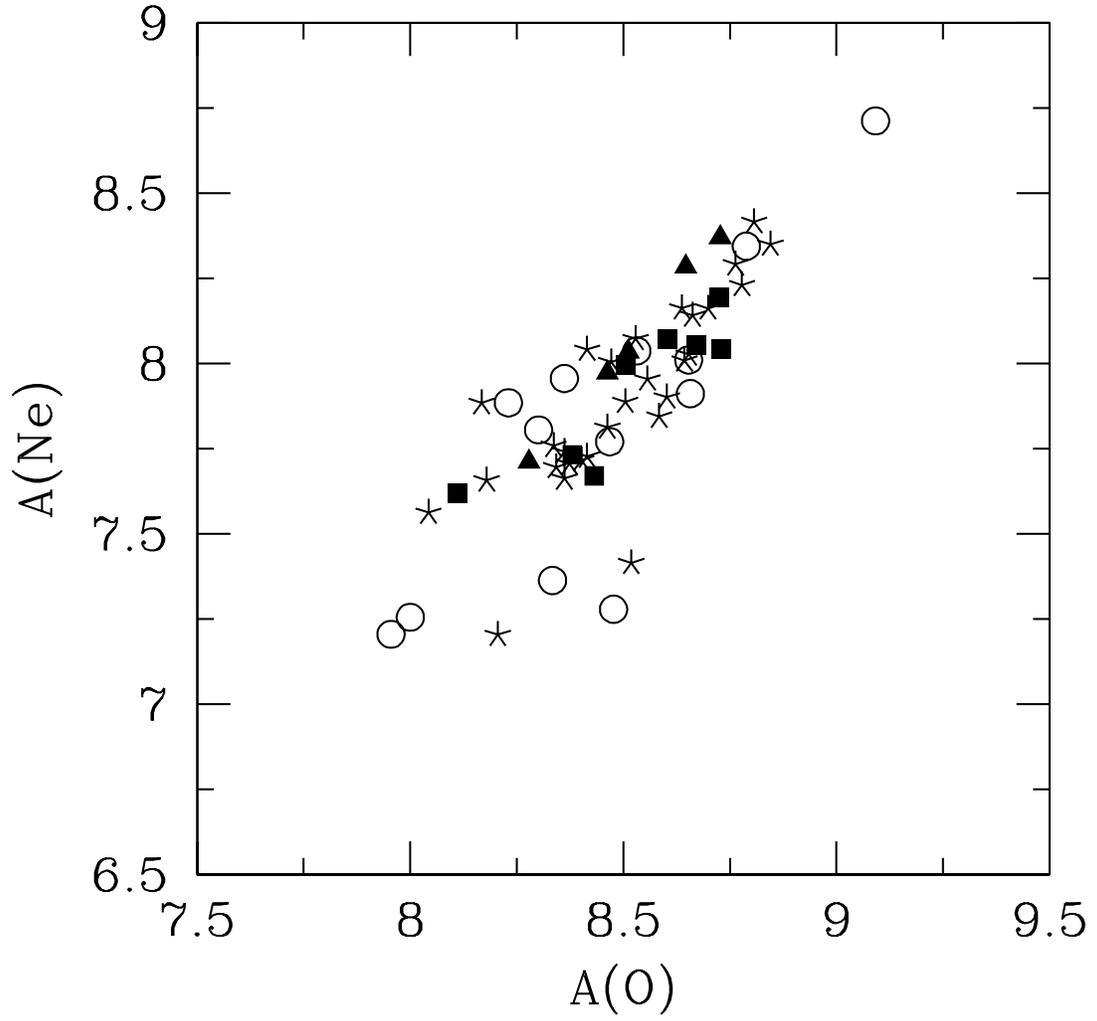}

\caption{A(Ne) versus A(O). Symbols are as in Figure 1.}

\end{figure}

\begin{figure}

\plotone{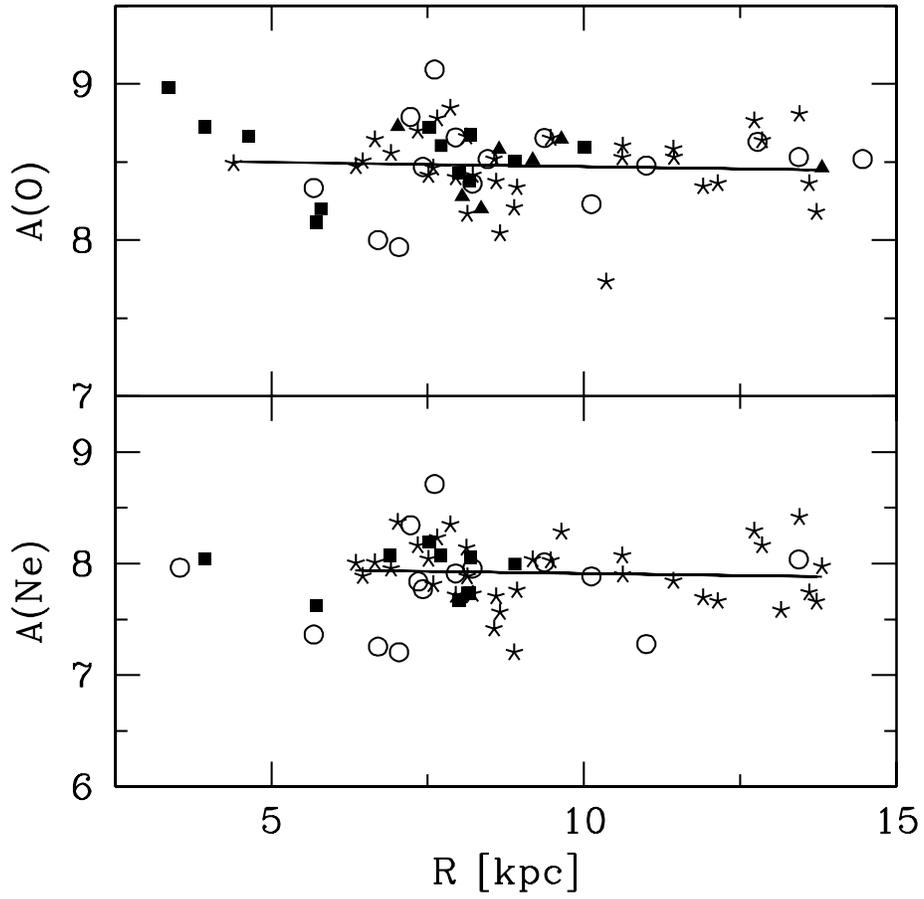}

\caption{Oxygen (top) and neon (bottom) abundances, given as log(X/H)+12,
versus galactocentric distances. Symbols are as in Figure 1.
The lines represent least squares fits, their inclination given in Table 3, and
they include E and BC PNe only.}

\end{figure}

\clearpage

\end{document}